\documentstyle[12pt]{article}
\newcounter{mycount}

\newcommand{\bee}{\begin{eqnarray}}
\newcommand{\eee}{\end{eqnarray}}
\newcommand{\be}{\begin{eqnarray}}
\newcommand{\ee}{\end{eqnarray}}

\newcommand\HHH{SH_3^\prime(0)}
\newcommand\I{{\cal I}}
\newcommand\C{${\open C}\left [ S_3 \right ]$}
\newcommand\A{{\cal A}}
\newcommand\nn{\nonumber \\}

\newcommand{\p}{\partial}

\font\frtnfr=eufm10   scaled\magstep1
\font\twlfr=eufm10
\font\tenfr=eufm10

\newfam\frfam
\textfont\frfam=\frtnfr
\scriptfont\frfam=\twlfr
\scriptscriptfont\frfam=\tenfr
\def\fr{\fam\frfam}

\font\frtnopen=msbm10  scaled\magstep2
\font\twlopen=msbm10
\font\tenopen=msbm10

\newfam\openfam
\textfont\openfam=\frtnopen
\scriptfont\openfam=\twlopen
\scriptscriptfont\openfam=\tenopen
\def\open{\fam\openfam}

\font\frtnsf = cmss12 scaled\magstep1
\font\twlsf = cmss10
\font\tensf = cmss9

\newfam\Scfam
\textfont\Scfam = \frtnsf
\scriptfont\Scfam = \twlsf
\scriptscriptfont\Scfam = \tensf

\begin{document}
\renewcommand{\theequation}{\arabic{equation}}
\bibliographystyle{nphys}
\setcounter{equation}{0}
%%%%%%%%%%%%%%%%%%%%%%%%%%%%%%%%%%%%%%%%%%%%%%
%%%%%%%%%%%%%%%%%%%%%%%%%%%%%%%%%%%%%%%%%%%%%%
\sloppy
\title
 {
      \hfill{\normalsize\sf math-ph/0112063}    \\
            \vspace{1cm}
            An example of simple Lie superalgebra
            with several invariant bilinear forms
 }
\author
 {
              S.E.Konstein
          \thanks
             {E-mail: konstein@lpi.ru}
          \thanks
             {
               This work was supported
               by the Russian Basic Research Foundation, grant 99-02-17916
               and grant 00-15-96566.
             }
  \\
               {\small \phantom{uuu}}
  \\
               {\sf \small I.E.Tamm Department of Theoretical Physics,}
  \\
               {\sf \small P. N. Lebedev Physical Institute,}
  \\
               {\sf \small 117924, Leninsky Prospect 53, Moscow, Russia.}
 }
\date{ }
%-----------------------------------------------------
\maketitle

\renewcommand{\theequation}{\arabic{equation}}
\setcounter{equation}{0}
%%%%%%%%%%%%%%%%%%%%%%%%%%%%%%%%%%%%%%%%%%%%%%%%%%%%%%
\begin{abstract}
{
%\footnotesize
Simple associative superalgebra with 2 independent supertraces is presented.
Its commutant is a simple
Lie superalgebra which has at least 2 independent invariant bilinear forms.
}
\end{abstract}

\newpage

\section{Introduction}

In this paper the associative superalgebra $\A=\HHH$ with
two-dimensional space of supertraces
is presented.
It is shown that
(i) it is simple,
(ii) its commutant ${\fr A}_1=\left[ \HHH, \, \HHH \right \}$
is a simple Lie superalgebra and
(iii) ${\fr A}_1$ has at least
2-dimensional space of nondegenerate bilinear invariant forms.
Would ${\fr A}_1$ be finitedimensional superalgebra having 2-dimensional
space of invariant bilinear forms, it necessarily would not be simple.

The superalgebra $\HHH$ belongs to the class of {\it superalgebras of
observables of $N$-particle Calogero model} which is denoted
in \cite{kv} as $SH^\prime_N (\nu)$, where $\nu$ is a coupling constant of
this Calogero model \cite{4}.

Let $x_i\in {\open R}$, $i=1,\, ... \,,  N$.

The Hamiltonian of Calogero model
after some similarity transformation attains the form
\be\label{Cal}
H_{Cal} =-\frac 1 2
\sum_{i=1}^N \left[ \frac {\p^2} {\p x_i^2} -x_i^2
+ \nu \sum_{j\neq i} \frac 2 {x_i - x_j}
\frac {\p} {\p x_i} \right]
\ee
which coincides with the operator
\be\label{Ca}
H=\frac 1 2 \sum_{i=1}^N \left\{a^0_i,\, a^1_i\right\}
\ee
on the space of symmetric functions.
The annihilation and creation operators
$a^\alpha_i$ ($\alpha=0,\,1$ correspondingly)
acting on the space of complex valued
infinitely smooth functions of $x_1$, $x_2$, ... , $x_N$ are expressed via the
differential-difference Dunkl operators  $D_i(x)$ \cite{6}
\bee\label{dunkl}
a^\alpha_i= \frac 1 {\sqrt{2}} (x_i +(-1)^\alpha D_i(x)),
\mbox { where }
D_i= \frac \p {\p x_i}+
\nu \sum_{l\neq i}^N  \frac {1} {x_i-x_l}(1-K_{il}),
\eee
and $K_{ij}=K_{ji}$ are the operators of elementary transpositions
\be
K_{ij} x_i = x_j K_{ij},\quad K_{ij}x_k=x_k K_{ij} \mbox{ if }k\neq i
\mbox{ and } k\neq j,
\ee
satisfying as a consequence the following relations:
\be
K_{ij} a^\alpha_i = a^\alpha_j K_{ij},\quad
K_{ij}a^\alpha_k=a^\alpha_k K_{ij} \mbox{ if }k\neq i
\mbox{ and } k\neq j,\ \ K_{ij}^2=1.
\ee

The operators  $a^\alpha_i$ satisfy the following commutation relations
\bee\label{AA}
\left [ a^\alpha_i\,,\,a^\beta_j \right ]= \epsilon^{\alpha\,\beta}
\left ( \delta_{ij} +
\nu\delta_{ij}\sum_{l=1,\,l\neq i}^N K_{il}-\nu\delta_{i\neq j}K_{ij}
\right),\nonumber
\eee
where
$\epsilon^{\alpha\,\beta}$ = $- \epsilon^{\beta\,\alpha},$
$\epsilon^{0\,1}=1$, which lead \cite{2} to the relations
\be
\left [H,\, a^\alpha_i \right ] = - (-1)^\alpha a^\alpha_i ,\nonumber
\ee
and, therefore, $a_i^1$ when acting on the Fock vacuum $|0\rangle$
such that $a_j^0|0\rangle = 0$ for any $j$,
give all the wavefunctions of the model (\ref{Ca})
and, consequently, of the model (\ref{Cal}). In \cite{3}
Perelomov operators \cite{Per},
describing the wavefunctions of the model (\ref{Cal}) are expressed
via these operators $a^\alpha_i$.

{\it {\bf Definition 1.} $SH_N(\nu)$ is
the associative superalgebra of all polynomials in the
operators  $K_{ij}$   and
$a^\alpha_i$
with
${\open Z}_2$-grading $\pi$ defined by the formula:
$\pi(a^\alpha_i)=1$, $\pi(K_{ij})=0$.}

The superalgebra $SH_N(\nu)$ is called
{\it the algebra of observables}.

It is clear that $SH_N(\nu) = SH_1(0)\otimes SH_N^\prime(\nu)$,
where $SH_1(0)$ is generated by the operators
$a^\alpha = \sum_i a^\alpha_i/\sqrt N$
satisfying the commutation relations
$[a^\alpha, a^\beta] =\epsilon^{\alpha\beta}$,
while $SH_N^\prime(\nu)$ is generated by the transpositions $K_{ij}$
and by all linear combinations of the form
$\sum_i \lambda_i a^\alpha_i$ with coefficients $\lambda_i$
satisfying the condition
$\sum_i\lambda_i=0$.

The group algebra
${\open C}[S_N]$ of the permutation group $S_N$,
generated by elementary
transpositions $K_{ij}$ is a  subalgebra of $SH_N^\prime(\nu)$.

{\it {\bf Definition 2.}
The supertrace on an arbitrary associative superalgebra $\A$
is  complex-valued linear function
$str(\cdot)$
on $\A$
which satisfies the condition
\be\label{str}
str(fg)=(-1)^{\pi(f)\pi(g)} str(gf)
\ee
for every $f,g \in \A$ with definite parity.}

It is proven in \cite{kv} that $SH_N(\nu)$  has nontrivial supertraces and
the dimension of the space of the supertraces on $SH_N(\nu)$
is equal to the number of partitions of $N$
into the sum of positive odd integers.
\footnote{The analogous result for Calogero model based on
arbitrary root system \cite{OP} is presented in \cite{F4}.}
In particular,
$SH_2(\nu)$
has only one supertrace and
$SH_3(\nu)$
has two independent supertraces. Since
$SH_N(\nu) = SH_1(0)\otimes SH_N^\prime(\nu)$
and $SH_1(0)$ has one supertrace, it follows that
the numbers of supertraces on $SH_N(\nu)$ and on $SH_N^\prime(\nu)$
are the same.

Every supertrace $str(\cdot)$ defines
the bilinear form $B_{str}$:
\be\label{form}
B_{str}(f,\,g)=str(fg).
\ee
Since the complete set of null-vectors
of every invariant bilinear form constitutes
the two-sided ideal of the algebra,
the supertrace is a good tool for finding
such ideals.

It should be noticed that the existence of supertraces
depends on the choice of
${\open Z}_2$-grading $\pi$,
while the existence of an ideals does not depend on the grading.

In the well known case of $SH_2^\prime(\nu)$
corresponding to the usual two-particle Calogero model
the only supertrace defines the bilinear form
(\ref{form}) which degenerates
when $\nu$ is half-integer
\cite{14}, hence, the superalgebra
$SH_2^\prime(\nu)$
has an ideal for these values of
$\nu$.
When
$N\ge 3$
the situation is more complicated
as $SH_N^\prime (\nu)$ has more than one supertrace.
In the case of finitedimensional
superalgebra this would be sufficient for existence of ideals
but for infinitedimensional superalgebras under consideration
the existence of ideals is not completely
investigated yet.

It is shown in \cite{N3}
that in the superalgebras
$SH_3^\prime (\nu)$ with
$\nu=n\pm 1/3$
or
$\nu=n+1/2$ ($n$ is an integer)
there exists one supertrace with degenerated
corresponding bilinear form (\ref{form}) and
that for the other values of $\nu$
all the bilinear forms corresponding to supertraces are non-degenerate.

One can suggest that for the latest case the corresponding superalgebras
are simple.

Below it is shown that the associative superalgebra
$SH_3^\prime (0)$ is simple and that its commutant is simple
Lie superalgebra.

The paper is organized as follows. In Section \ref{algebra}
the generating elements of associative superalgebra
$SH_3^\prime(0)$
are presented together with the relations between them.
The simplicity of
$SH_3^\prime(0)$ is proved in Section \ref{simple} and in the last section
the simplicity of $[SH_3^\prime(0),\, SH_3^\prime(0)\}$
is established.

\section{The algebra $SH_3^\prime(0)$}\label{algebra}
In this Section we present the generating elements of
associative superalgebra
$\A=SH_3^\prime(0)$ and relations between them.

Let
$\lambda = exp(2\pi i/3)$.
       As a basis in
${\open C}\left [ S_3 \right ]$ let us choose
the vectors
\bee
L_k &=&  \frac 1 3 (\lambda^k K_{12}+K_{23}+\lambda^{-k} K_{31}), \nn
Q_k &=& \frac 1 3 (1 + \lambda^k K_{12}K_{13} + \lambda^{-k} K_{12}K_{23}),\nn
&{}&L_{i+3}=L_i,\quad Q_{i+3}=Q_i, \nn
&{}&L_\pm\stackrel {def} {=}  L_{\pm 1},\quad
Q_\pm \stackrel {def} {=} Q_{\pm 1}. \nonumber
\eee
Instead of the generating elements
$a^\alpha_i$ let us introduce
the vectors
\bee
x^\alpha &=& a^\alpha_1+ \lambda a^\alpha_2+ \lambda^2 a^\alpha_3,
\ \ \ x \stackrel {def} {=} x^0, \ \ \ x^+ \stackrel {def} {=} x^1,   \nn
y^\alpha &=& a^\alpha_1+ \lambda^2 a^\alpha_2+ \lambda a^\alpha_3,
\ \ \ y \stackrel {def} {=} y^0, \ \ \ y^+ \stackrel {def} {=} y^1,  \nonumber
\eee
which are derived from
$a^\alpha_i$
by subtracting the center of mass
$a^\alpha =
\displaystyle{\frac 1 3 \sum_{i=1,2,3} a^\alpha_i}$.

The Lie algebra
${\fr sl}_2$
of inner automorphisms of
$SH_3^\prime(0)$
is generated by the generators
$T^{\alpha \beta}$
\bee\label{sl2}
T^{\alpha \beta} = \frac 1 3 (x^\alpha y^\beta + x^\beta y^\alpha).
\ee
These generators satisfy the usual commutation relations
\be \label {csl2}
        [T^{\alpha\beta}, T^{\gamma\delta}] =
   \epsilon^{\alpha\gamma} T^{\beta\delta}  +
   \epsilon^{\alpha\delta} T^{\beta\gamma}  +
   \epsilon^{\beta\gamma}  T^{\alpha\delta} +
   \epsilon^{\beta\delta}  T^{\alpha\gamma} \,,
\ee
and act on generating elements
$x^\alpha$
and
$y^\alpha$
as follows:
\be \label {sl2vec}
\left [ T^{\alpha\beta},\,x^\gamma \right ]=\epsilon^{\alpha\gamma} x^\beta
                                  +\epsilon^{\beta\gamma} x^\alpha \,,\qquad
\left [ T^{\alpha\beta},\,y^\gamma \right ]=\epsilon^{\alpha\gamma} y^\beta
                                  +\epsilon^{\beta\gamma} y^\alpha
\ee
leaving the group algebra ${\open C}[S_3]$ invariant:
$\left[ T^{\alpha\beta},\, K_{ij}\right]=0$.

Clearly, $SH_3^\prime(0)$
decomposes into the infinite direct sum of finitedimensional
irreducible representations of this ${\fr sl}_2$.
Let ${\cal A}_s$ ($s$ is a spin)
be the direct sum of all $2s+1$-dimensional irreducible
representations in this decomposition.
So
\be
\A={\cal A}_0\oplus {\cal A}_{1/2} \oplus {\cal A}_1 \oplus \,...
\ee
Evidently, ${\cal A}_0$ is a subalgebra of $\A=SH_3^\prime(0)$.

As every element of the subspace ${\cal A}_s$ for $s>0$
is a linear combinations of the vectors of the form
$[f,\,T^{\alpha\beta}]$ where $T^{\alpha\beta}$ is defined in (\ref{sl2}),
every supertrace vanishes on this
subspace and has non-trivial values only on the associative
subalgebra ${\cal A}_0 \subset SH_3^\prime(0)$ of all ${\fr sl}_2$-singlets.

Let
$m=\frac 1 4 \left \{ x^\alpha,\, y_\alpha\right\}$.
Clearly, $m$ is a singlet under
the              action of ${\fr sl}_2$
(\ref{sl2});
$m$ can also be expressed in the form
\bee
\label{mx}
m &=& \frac 1 2 \left  ( x^\alpha y_\alpha +3
\right)
\mbox{   or, equivalently,}\\
\label{my}
m &=& \frac 1 2 \left  ( y_\alpha x^\alpha -3
\right).
\eee
In these formulas the greek indices are lowered and raised
with the help of the antisymmetric tensor $\varepsilon^{\alpha\beta}$:
$a^{\alpha}= \sum_\beta \varepsilon^{\alpha\beta}a_\beta $.

The generating elements
$x^\alpha$,
$y^\alpha$,
$Q_i$,
$L_i$ and the element $m$
satisfy the following relations:
%%%%%%%%%%%%%%%%%%%%%%%
%  $S_3$ and $S_3$:
%%%%%%%%%%%%%%%%%%%%%%%
\bee
\label{LL}
L_i L_j = \delta_{i+j} Q_j , \qquad
\label{LQ}
L_i Q_j = \delta_{i-j} L_j , \\
\label{QL}
Q_i L_j = \delta_{i+j} L_j ,  \qquad
\label{QQ}
Q_i Q_j = \delta_{i-j} Q_j ,
\eee
where $\delta_0=1$ and $\delta_i=0$ if $i \neq 0$,
%%%%%%%%%%%%%%%%%%%%%%%
%  $S_3$ and $SH_2$:   %
%%%%%%%%%%%%%%%%%%%%%%%
\bee
\label{Lx}
L_i x^\alpha = y^\alpha L_{i+1},  & \qquad &
\label{Ly}
L_i y^\alpha = x^\alpha L_{i-1},   \\
\label{Qx}
Q_i x^\alpha = x^\alpha Q_{i+1},  & \qquad &
\label{Qy}
Q_i y^\alpha = y^\alpha Q_{i-1},   \\
\label{Lm}
L_i m = -m L_i                 ,  & \qquad &
\label{Qm}
Q_i m =  m Q_i                 ,
\eee
%%%%%%%%%%%%%%%%%%%%%%%%%%
%  $SH_2$ and $SH_2$:  %%%%%
%%%%%%%%%%%%%%%%%%%%%%%%%%
\bee
\left [ x, \, x^+ \right ]  =
0
,\qquad
\left [ y, \, y^+  \right ] =
0
,\qquad
\left [ y, \, x^+  \right ] =\left [ x, \, y^+ \right ] = 3
,\nn
%\qquad
\left [ x, \, y  \right ] =\left [ x^+, \, y^+ \right ] = 0
,\eee
and
\bee\label{MY}
\left [ m, \, x^\alpha  \right ] &=&\frac 3 2 \left ( x^\alpha
\right ),         \nn
\left [ m, \, y^\alpha \right ] &=& - \frac 3 2 \left ( y^\alpha
\right ),
\eee
%from which the relations
%%%%%%%%%%%%%%%%%%%%%%%%%%%
%%  Consequences:     %%%%%
%%%%%%%%%%%%%%%%%%%%%%%%%%%
%\bee
%\label{mx-}
%mQ_+ x^\alpha Q_- &=& Q_+ x^\alpha Q_- \left (m+\frac 3 2 \right ) ,   \\
%\label{my+}
%mQ_- y^\alpha Q_+ &=& Q_- y^\alpha Q_+ \left (m-\frac 3 2 \right ) ,   \\
%\label{mx0}
%mQ_- x^\alpha Q_0 &=& Q_- x^\alpha Q_0 \left (m+\frac 3 2
%\right )   ,   \\
%\label{my0}
%mQ_+ y^\alpha Q_0 &=& Q_+ y^\alpha Q_0 \left (m-\frac 3 2
%\right )   ,   \\
%Q_0 x^\alpha Q_+ m &=& \left (m-\frac 3 2
%\right ) Q_0 x^\alpha Q_+   ,   \\
%\label{ym-}
%Q_0 y^\alpha Q_- m &=&  \left (m+\frac 3 2
%\right ) Q_0 y^\alpha Q_-
%\eee
%follow.

{\it Corollary 1.} There exists ${\open Z}_3$ grading $\rho$
on $SH_3^\prime(0)$ defined by the formulas:

$\rho (Q_i) =0$,

$\rho (L_i) = -i$,

$\rho (x^\alpha) =1$,

$\rho (y^\alpha) = -1$,\\
such that $fQ_i = Q_if$ for any element $f\in \HHH$ with $\rho (f)=0$.
Particularly, $[(x^\alpha)^3,\,Q_i] =[(y^\alpha)^3,\,Q_i] =0$.

{\it Corollary 2.} There exists the antiautomorphism $\tau$
of associative algebra $\HHH$:\\
$\tau (x^\alpha)=y^{1-\alpha}$, $\tau (y^\alpha)=x^{1-\alpha}$,  $\tau(m)=m$,\\
$\tau (L_i) = L_{-i}$, $\tau (Q_i)=Q_i$.

Every supertrace on
$SH_3^\prime(\nu)$
is completely determined
by its values on
${\open C}\left [ S_3 \right ] \subset SH_3^\prime(\nu)$,
 i.e.,   by the values
$str(1)$, $str(K_{12})$ = $str(K_{23})$ = $str(K_{31})$
and
$str(K_{12}K_{23}) = str(K_{12}K_{13})$,
which are consistent with (\ref{str}) and
admit the extension of the supertrace from
${\open C}\left [ S_3 \right ]$
to
$SH_3^\prime(\nu)$ in a unique way, if and only if
$
str(K_{ij})= \nu (-2str(1) - str(K_{12}K_{23}))
$ \cite{kv}.

It was noticed above that
only trivial
representations of ${\fr sl}_2$ can contribute to any supertrace on
$SH_3^\prime(0)$.
Evidently, the associative algebra
${\cal A}_0$
of ${\fr sl}_2$-singlets contains only
the polynomials of $m$ with coefficients
in \C.

To describe the restrictions of all the supertraces on ${\cal A}_0$,
it is convenient to use the generating functions, which are
computed in \cite{N3} for any values of $\nu$, and for $\nu=0$
take the following form:
\bee
{\cal L}_i &\stackrel {def} {=} & str (e^{2/3\xi m} L_{i})=0 ,\\
\label{genfun}
    {\cal Q}_i & \stackrel {def} {=} & str (e^{2/3\xi m} Q_{i}) =
    \frac {P_i}{\Delta},
\eee
    where
\bee
%%%%%%%%%%%%%%%%%%%%%%%%%%%%%%%%%%%%%%%%%%%%%%%%%%%%%%%%%%%%%%%%%%%%
      P_0 &=& -2 S_1
     + \frac {S_2} {2}
     \left( e^{2\xi} + e^{-2\xi} - 2 e^{\xi} -2 e^{-\xi}  \right),\\
%%%%%%%%%%%%%%%%%%%%%%%%%%%%%%%%%%%%%%%%%%%%%%%%%%%%%%%%%%%%%%%%%%%%
    P_+ &=&
     {2\over 3} S_1 \left( -e^{ 2\xi} + 2 e^{- \xi}\right)
       +
 \frac {S_2} {2} \left( e^{- 2\xi} -2 e^{ \xi} + 3 \right),\\
%%%%%%%%%%%%%%%%%%%%%%%%%%%%%%%%%%%%%%%%%%%%%%%%%%%%%%%%%%%%%%%%%%%%
    P_- &=&
     {2\over 3} S_1
      \left( -e^{-2\xi} + 2 e^{\xi}\right)
     + \frac {S_2} {2}
     \left( e^{ 2\xi} -2 e^{-\xi} + 3 \right)\\
%%%%%%%%%%%%%%%%%%%%%%%%%%%%%%%%%%%%%%%%%%%%%%%%%%%%%%%%%%%%%%%%%%%%
\Delta &=& exp(-3\xi)(exp(3\xi)+1)^2.
\eee
Here
$S_1$, $S_2$  are arbitrary parameters, specifying the supertrace
in the two-dimensional space
of supertraces:
\bee
S_1=-2str(1)-str(K_{12}K_{23}),\qquad
S_2=\frac 8 3 str(1) -\frac 2 3 str(K_{12}K_{23}) .  \nonumber
\eee

\section{The simplicity of $\HHH$.}\label{simple}

To prove the simplicity of $\HHH$ it is sufficient to prove that elements
$Q_i$ belong to any nonzero ideal $\I$.

Let $f\in \I$, $f=f_0+f_{1/2}+f_1+...$,
where $f_s\in{\cal A}_s$. As $[T^{\alpha\beta},\,f]\in\I$ then
$f_s\in \I$. So each ideal in ${\cal A}$ is graded.

{\bf Proposition 1.}
{\it Let $\I\subset \HHH$ be two-sided nonzero ideal. Then
there exist nonzero polynomial $f(x^\alpha,\,y^\alpha)$ and a polynomial
$g(x^\alpha,\,y^\alpha)$
such that $fQ_0 + g L_0\in \I$}.

Indeed, let $r\in \I$, $r\neq 0$. Then at least one of the polynomials
$rQ_0$, $rL_0$, $rxQ_0$, $rxL_0$, $ryQ_0$, $ryL_0$ has the desired form.

{\bf Proposition 2.}
{\it Let $\I\subset \HHH$ be two-sided nonzero ideal. Then
there exist nonzero polynomial $f_0(m)$ and a polynomial
$g_0(m)$
such that $f_0Q_0 + g_0 L_0\in \I$}.

Let $f(x^\alpha,\,y^\alpha)\neq 0$
and $h_0=fQ_0 + g L_0\in \I\bigcap \A_s$ with some polynomial
$g(x^\alpha,\,y^\alpha)$. Let
\bee
h_{k+1}=\left[T^{00},\,h_k\right]
\eee
Evidently, $h_k \in \I$ for any $k$.
Then there exists such integer $K$ that $h_K\neq 0$
and $h_{K+1}=0$. The vector $h_K$ is the highest vector of some
irreducible representation of ${\fr sl}_2$ algebra (\ref{sl2})
and has the form
\bee
h_K=\sum_{i=0}^{2s} \left((x)^i(y)^{2s-i}q_i(m)Q_0
+(x)^i(y)^{2s-i}p_i(m)L_0 \right)
\eee
where some polynomial $q_n(m)\neq 0$.
Then $F=(y^+)^n(x^+)^{2s-n}h_K \in \I$.
Let $F_0$ be the singlet part of $F$. Evidently, $F_0\neq 0$,
$F_0\in \I$, and $F_0=f_0(m)Q_0+g_0(m)L_0$ with nonzero polynomial
$f_0(m)$.

{\bf Proposition 3.}
{\it Let $\I\subset \HHH$ be two-sided nonzero ideal. Then
there exists nonzero polynomial $f_0(m)$
such that $F_0=f_0(m)Q_0 \in \I$}.

Let $F=f(m)Q_0+g(m)L_0\in \I$, $f(m)\neq 0$.
Then $mf(m)Q_0=\frac 1  2 (mF+Fm)\in\I$ due to commutational relations.

{\bf Proposition 4.}
{\it Let $\I\subset \HHH$ be two-sided nonzero ideal. Then
there exist nonzero polynomials $f_i(m)$
such that $F_i=f_i(m)Q_i \in \I$} for $i=0,\, \pm 1$.

Let $F_0=f_0(m)Q_0\in \I$ is nonzero polynomial from Proposition 3.
Let us denote the singlet part of arbitrary element $F\in \A$ as $(F)_0$
Then $F_1=(y^+F_0x)_0=f_1(m)Q_1$  and $F_2=(x^+F_0y)_0=f_2(m)Q_2$
are nonzero and belong to $\I$.

{\bf Proposition 5.}
{\it Let $\I_i=\{f(m)\in {\open C}[m]:\,\, f(m)Q_i\in\I\}$.
Then $\I_i$ are nonzero ideals in ${\open C}[m]$}.

This proposition follows trivially from Proposition 4.

Hence there exist polynomials $\varphi_i(m)=m^{k_i}+...$ such that
$\I_i=\varphi_i (m)\cdot {\open C}[m]$.

{\bf Proposition 6.} {\it The polynomials $\varphi_i(m)$ generating the
ideals $\I_i$ satisfy the relations}
\be
\varphi_0 (m)=\pm \varphi_0 (-m),\\
\varphi_1 (m)=\pm \varphi_2 (-m).
\ee
Indeed, $\varphi_0(-m)Q_0=L_0 \varphi_0(m) Q_0L_0 \in \I$,
hence $\varphi_0(-m)\in \I_0$ and
 all the roots of $\varphi_0(m)$ are the roots of $\varphi_0(-m)$.
{}Further, $L_1\varphi_1(m)Q_1L_2=\varphi_1(-m) Q_2$ and, as a consequence,
$\varphi_1(-m)=p(m) \varphi_2(m)$ with some nonzero polynomial $p(m)$.
Analogously, $\varphi_2(-m)=q(m) \varphi_1(m)$
with some nonzero polynomial $q(m)$. So $p(m)q(-m)=1$,
and $\varphi_2(-m)=\pm \varphi_1(m)$.

Now we can prove the following theorem:

{\bf Theorem 1.} {\it The associative superalgebra $\A=\HHH$ is simple.}

We will prove that arbitrary nonzero ideal in $\A$ contains
the element $1\in\A$.
Let $\I$ be nonzero ideal in $\A$ and $\varphi_i(m)$ generate the ideals
$\I_i$ described above.
So $\varphi_i(m)Q_i\in\I$ and hence
$y^3\varphi_i(m)Q_i(x^+)^3\in\I$.
Consider the ${\fr sl}_2$-singlet part
of this element which also belongs to $\I$ and can be computed with the help
of (\ref{MY}):
\be
(y^3\varphi_i(m)Q_i(x^+)^3)_0
=P(m)\varphi_i(m+9/2) Q_i
\in\I,\\
\mbox{where } P(m)=const(m+3/2)(m+3)(m+9/2)\neq 0.
\ee
So $
P(m)\varphi_i (m+9/2)\in \I_i$
and
$$
P(m)\varphi_i (m+9/2)=p_i(m)\varphi_i (m)
$$
with some polynomials $p_i(m)$. Since $\varphi_0(m)$ is even or odd polynomial,
the real part of the rightest root of $\varphi_0(m)$
is non negative, and so
this rightest root is not a root of polynomial
$(m+3/2)(m+3)(m+9/2)\varphi_0 (m+3/2)$.
So $\varphi_0(m)$ has no roots and $\varphi_0(m)=1$.
One can apply the same consideration to the function $\varphi(m)=
\varphi_1(m)\varphi_2(m)$
which is even due to Proposition 6 and satisfies the equation
$$
P(m)^2\varphi (m+9/2)=p_1(m)p_2(m)\varphi (m).
$$
So $\varphi_1(m)\varphi_2(m)=1$ and $\varphi_1(m)=\varphi_2(m)=1$.

In such a way, $Q_i\in \I$ and $1=Q_0+Q_1+Q_2\in\I$.

%%%%%%%%%%%%%%%%%%%%%%%%%%%%%%%%%%%%%%%%%%%%%%%%%%%%%%%%%%

\section{Lie superalgebra $\A^L$}\label{last}

Now let us consider Lie superalgebra ${\fr A}=\A^L$ which consists of elements
of associative algebra $\A=\HHH$ with operation
$[f,g\}=fg - (-1)^{\pi(f)\pi(g)}gf$.
This Lie superalgebra has an ideal ${\fr A}_1=[{\fr A},\,{\fr A}\}$
with the following evident properties

(i) $codim {\fr A}_1 =2$,

(ii) ${\cal A}_{1/2} \oplus {\cal A}_1 \oplus \,...\subset {\fr A}_1 $.

Let $Z$ be the center of $\HHH$. Obviously, $Z\subset {\cal A}_0$
and so every element of $Z$ is a polynomial of $m$ with coefficients
in \C.

{\bf Proposition 7.} $Z={\open C}$.

Let $f(m)\in Z$. As $m$ is even then $f$ is even element also.
As $mL_i=-L_im$, the relation $[f,\,m]=0$ gives
that $f(m)=\sum f_i(m)Q_i$.
Due to equalities $[(x)^3,\,Q_i]=0$ and (\ref{MY})
one has
$[(x)^3,\,f(m)]=(x)^3(f(m)-f(m-9/2))$ and as a consequence
$f(m)=f(m-9/2)$. So $f(m)$ does not depend on $m$, $f(m)\in$\C.
To finish the proof it is  sufficient to commute $f$ with $x$ or $y$.

{\bf Proposition 8.} $Z\bigcap {\fr A}_1 = \{0\}$.

As there exists such supertrace that $str(1)\neq 0$ and $Z={\open C}$,
it follows from  $str(Z\bigcap {\fr A}_1)=0$ that
$Z\bigcap {\fr A}_1 = \{0\}$.

{\bf Proposition 9.}
{\it Let $\I\subset {\fr A}_1$ be an ideal and $\I_{even}=\{0\}$.
Then $\I_{odd}=\{0\}$.}

Let $r\in\I_{odd}$.
Due to ${\fr sl}_2$ invariance one can choose
$r$ in the form
$$
r=\sum_{i=0,1,2}\sum_{n=0}^{2s}
\left( (x)^n(y)^{2s-n} f_{ni}(m)Q_i
+(x)^n(y)^{2s-n} g_{ni}(m)L_i \right)
$$
with odd value of $2s$.
As $(x)^3\in \A_{3/2}\subset {\fr A}_1$ then
$[(x)^3,\, r\}\in \I_{even}$ and $\{(x)^3,\, r\}=0$.
So the following relations
take place:
\bee
\sum_{i=0,1,2}\sum_{n=0}^{2s}
(x)^{n+3}(y)^{2s-n} \left(f_{ni}(m)+f_{ni}(m+9/2)\right)Q_i=0
\eee
and so $f_{ni}=0$.

{}Further, it follows from
$0=\{(x)^3,\, r\}\simeq
\sum_{i=0,1,2}\sum_{n=0}^{2s}
\left((x)^3+(y)^3\right) (x)^n(y)^{2s-n} g_{ni}(m)L_i$
that $g_ni(m)=0$ also.
Here the sign $\simeq$ is used to denote
the equality up to polynomials of lesser degrees.

{\bf Theorem 2} {\it Lie superalgebra ${\fr A}_1$ is simple.}

This Theorem follows from the next two general theorems proven by
S.Montgomery in \cite{SM}.

{\bf Theorem 3} {\it Let $\A$ be associative simple superalgebra and
$\I\subset [\A^L,\,\A^L\}$ be a graded ideal of $[\A^L,\,\A^L\}$ such that
$\I\neq [\A^L,\,\A^L\}$. Then $\I_3=\{0\}$, where $\I_1=[\I,\,\I\}$,
$\I_2=[\I_1,\,\I_1\}$, $\I_3=[\I_2,\,\I_2\}$.}

{\bf Theorem 4.} {\it Let $\A$ be associative simple superalgebra and
$\I\subset [\A^L,\,\A^L\}$ be agraded ideal of $[\A^L,\,\A^L\}$ such that
$[\I_{even},\,\I\} \subseteq Z(\A)$ where $Z(\A)$ is the center of $\A$.
Then $\I_{even}\subseteq Z(\A)$.}

Indeed, let $\I\subset {\fr A}_1$ be an ideal such that $\I\neq {\fr A}_1$.
Due to Theorem 3 one can consider that $[\I,\,\I\}=\{0\}$.
Then it follows from Theorem 4 and Proposition 8 that $\I_{even}=\{0\}$.
{}Further, Proposition 9 gives that $\I_{odd}=\{0\}$, and so
$\I=\{0\}$.

{\bf Theorem 5} {\it Simple Lie superalgebra ${\fr A}_1$ has at least 2
independent bilinear invariant forms.}

Two-dimensional space of supertraces $str(\cdot)$ on $\A=\HHH$ generates some
space of invariant bilinear forms $B(u,\,v)=str(uv)$ on ${\fr A}_1$.
This space is also 2-dimensional.
Indeed, let some supertrace $str_0(\cdot)$
on $\A$ leads to bilinear
form $B_0$ on ${\fr A}_1$ which is equal to zero identically.
The elements $x$, $y^+$,
$xQ_1$ and $y^+Q_1$ belong to
${\cal A}_{1/2}\subset {\fr A}_1 $.
So, $0=B_0(x,y^+)-B_0(y^+,x)=str_0([x,\,y^+])=3str_0(1)$ and
$0=B_0(x,y^+Q_1)-B_0(y^+,xQ_1)=str_0([x,\,y^+]Q_1)=3str_0(Q_1)$.
It follows from (\ref{genfun}) that
$str_0(1)=-\frac 1 6 (S_1 - \frac 3 2 S_2)$ and
$str_0(Q_1)=\frac 1 6 (S_1 + \frac 3 2 S_2)$.
So $S_1=S_2=0$ and $str_0$ is equal to zero identically.

\vskip 5 mm
\vskip 5 mm
\noindent {\bf Acknowledgments} \vskip 3 mm
\noindent
The author is
very grateful to M.~Vasiliev and to D.~Leites for useful discussions.

\end{document}